\documentstyle[12pt]{article}

\newcommand{\pagenumber}{\pagestyle{plain}\setcounter{page}{1}}

\newcommand{\zi}{z_{i}}
\newcommand{\zj}{z_{j}}
\newcommand{\ki}{k_{i}}

\newcommand{\dktilde}{d\tilde{k}}
\newcommand{\dkitilde}{d\tilde{k}_{i}}
\newcommand{\Ki}{K_{i}}

\newcommand{\dx}{dx_{0}}

\newcommand{\dy}{dy_{0}}
\newcommand{\prop}[2]{G'(\sigma_{#1}, \sigma_{#2})}
\newcommand{\msr}[2]{\prod_{#1}^{#2} d^{2}z_{i}}
\newcommand{\pnorm}[1]{\prod_{#1} |\zi-\zj|^{2K_{i}\cdot K_{j}} }
\newcommand{\expo}[2]{2K_{#1}\cdot K_{#2}}
\newcommand{\Ti}{T_{i}}

\def\fnote#1#2{\begingroup\def\thefootnote{#1}\footnote{#2}
    \addtocounter{footnote}{-1}\endgroup}
\def\sppt{makoto@sbitp.ucsb.edu}

\begin{document}

\pagestyle{empty}

\begin{flushright}
    NSF-ITP-93-157\\
    UTTG-33-93\\
\end{flushright}
\vspace{24pt}
\begin{center}
{\bf Zero Mode Divergence Problem in String Theory}

\vspace{36pt}
Makoto Natsuume\fnote{*}{\sppt}

\vspace{6pt}
{\sl Institute for Theoretical Physics \\
University of California \\
Santa Barbara, California 93106-4030 }
\vspace{48pt}

\underline{ABSTRACT}

\end{center}

\vspace{24pt}
For $2D$ string theory,
the perturbative $S$-matrices are not well-defined
due to a zero mode divergence.
Although there exist formal procedures to make the integral
convergent, their physical meanings are unclear.
We describe a method to obtain finite $S$-matrices physically
to justify the formal schemes.
The scheme uses asymptotic states
by wave packets which fall faster than exponentials.
It is shown that the scheme gives well-defined $S$-matrices
and justifies the formal shifted Virasoro-Shapiro amplitude
for simple processes.
The tree-level unitarity for these processes is also shown.
We point out a problem in this scheme.
\vfill

\newpage
\pagenumber
\baselineskip=20pt 

\section{Introduction}

In this paper, we examine a method to obtain
well-defined $S$-matrix for $2D$ string, {\it i.e.}
$2D$ critical string theory propagating in a linear dilaton background
\cite{myers,ginsparg}.
In spite of the importance of the background,
$S$-matrix calculations appearing in the literature \cite{gross}
are formal, without dealing with the zero mode divergence properly;
thus, their amplitudes are really ill-defined.

The background is described by a nonlinear $\sigma$-model
and has the following world-sheet action
in two dimensional Minkowskian spacetime
\footnote{
Although its Euclidean version is commonly studied,
we will take the Minkowski signature
for the better physical interpretation, {\it i.e.}
physical scattering in two dimensional spacetime.
}
\begin{equation}
S = \frac{1}{8\pi} \int d^{2}\sigma \sqrt{g}
       \{ g^{ab} G_{\mu \nu}(X)
          \partial_{a} X^{\mu} \partial_{b} X^{\nu}
          + 2\Phi(X) R
          + T(X) \},
\label{eq:action}
\end{equation}
where $X^{\mu}(\sigma)=\left(t(\sigma), \phi(\sigma)\right)$.
$G_{\mu \nu}, \Phi$, and $T$ denote
the spacetime metric, the dilaton, and the tachyon respectively.
The background value for the fields are
$\bar{G}_{\mu \nu} = \eta_{\mu \nu}$,
$\bar{\Phi} = n^{\mu}X_{\mu}$, and $\bar{T} = 0$.
Conformal invariance for the background
restricts the value of $n$ such that $n^{2}=2$.
In order to avoid complex world-sheet action,
we take $n^{\mu}$ in spatial direction,
namely $n\!\cdot\! X = \sqrt{2}\phi$
(In this paper, we will not use this explicit form
of $n^{\mu}$ for simplicity.).

This background provides a very useful laboratory in string theory.
The background is particularly important
since the theory is one of the few known
exact conformal invariant backgrounds \cite{myers}.
Also,
the theory can be viewed as quantum gravity
(Liouville theory with the zero cosmological constant)
coupled to $c=1$ conformal matter
after the continuation to Euclidean spacetime metric \cite{ginsparg}.

However, the $S$-matrix
\footnote{
This is the tachyon $S$-matrix,
since the tachyon is the only propagating mode of $2D$ strings.
}
for the theory and its unitarity are not
well understood owing to the zero mode divergence
in the $S$-matrix calculations.
The standard path integral evaluation starts by
separating $X$
into the zero mode of the laplacian operator and the other eigenmodes,
which reduces the path integral to an infinite number
of ordinary integrals.
For the bosonic string,
the zero mode integral gives a momentum delta function;
whereas here, as we shall see in Section 3,
the integral actually diverges:
\begin{equation}
\int \prod_{\mu} dx_{0}^{\mu} \,
   e^{i c x_{0} \cdot \{\sum_{i=1}^{M}k_{i}-i(M-\chi)n \} },
\label{eq:zero_mode}
\end{equation}
where $\chi$ is the world-sheet Euler number
and $M$ is the number of vertex operators inserted.
The additional term in the exponent
appears because of the failure of spacetime translational invariance
for the action (\ref{eq:action}).

Physically, this zero mode divergence is
due to string coupling divergence.
Since the string coupling is given by $e^{\bar{\Phi}}$,
the coupling diverges for the linear background, $\Phi=\sqrt{2}\phi$,
at spatial infinity, $\phi \rightarrow \infty$.
As a result of the coupling divergence,
perturbative $S$-matrices based on plane wave asymptotic states
are not well defined,
since the plane waves couple at infinity where the coupling diverges.

There are various ways proposed in the literature
to obtain well-defined $S$-matrices.
The original suggestion was
to ignore the divergence and write the zero mode integral
formally as a delta function whose argument is complex \cite{myers}.
For the Liouville theory with Euclidean spacetime metric,
the standard procedure is the Wick rotation
of the Liouville mode by $\phi\! \rightarrow\! i\phi$,
which makes the integrand (\ref{eq:zero_mode})
oscillatory \cite{gross,minic};
this is Minkowski continuation of the spacetime metric,
taking $n^{\mu}$ in time direction instead of spatial direction.
The theory basically reduces to
the Feigen-Fuchs construction \cite{feigen,dotsenko}.

It is certainly welcome to have well-defined amplitudes,
but the physical meaning behind those formal regularizations
is not clear.
One is tempted to ask
what happens physically in the process of regularizations.
In other words, is it possible to justify the formal schemes?
Or, is there any physical scheme to obtain well-defined $S$-matrices?
It is natural to expect affirmative answers to the questions,
since analyses imply the consistency of the formal $S$-matrices;
for instance, tree-level unitarity is formally shown
by factorization analysis \cite{minic,sakai}.

In order to answer the above puzzle,
we propose a physical scheme for the reguralization
and compare the result with the formal one to justify the latter.
The scheme is based on localized asymptotic states
instead of plane waves.
Since the problem essentially comes from the fact that
the plane wave couples at infinity,
$S$-matrices are expected to be well-defined
if one uses asymptotic states
which are well-localized and peaked in the weakly coupled region.
\footnote{This was first pointed out by Polchinski.
See \cite{minic}.}

The organization of the present paper is as follows.
First, in the next section, we will calculate the $S$-matrices of
a toy field theoretical model, which is
the low energy effective theory derived from a nonlinear $\sigma$-model.
The aim here is to show the wave packet scheme explicitly
in a simpler context than the corresponding string calculation.
In Section 3, the corresponding string $S$-matrix calculations
will be carried out.
As a by-product of these calculations, unitarity will be shown explicitly
for simple amplitudes, where only formal proofs are known.
Then, in Section 4, we will point out that the wave packet scheme
does not always work.
In this paper, only tree-level amplitudes are subject to study,
because the wave packet scheme does not always work
already at the tree-level.

\section{Toy Model Calculation}

The asymptotic states used in the following sections
are represented as
superpositions of plane waves;
\begin{equation}
|i \rangle = \int \dkitilde\, \rho_{i}(\vec{\ki}) | \ki \rangle,
                                            \label{eq:wavepacket}
\end{equation}
where $\dkitilde = dp_{i}/(2\pi)(2\omega_{i})$.
The wave function is given by $T_{i}(X) = \langle X|i \rangle$.
We do not specify the explicit form of the wave packets;
the only requirement we have to impose is
that the wave function falls off faster than
any exponentials at spatial infinity, $\phi \rightarrow \infty$.

One important point to keep in mind is that
the scheme does not make {\it every} $S$-matrix well-defined.
The scheme is essentially a Hilbert space relabeling from
momentum $k$ to the index $i$.
Since asymptotic states are assumed to form a complete set
(asymptotic completeness),
the choice of the set should not affect
the behavior of $S$-matrices on the whole.
This implies that there should always exist divergent $S$-matrices
and the divergence is unavoidable
even by the use of wave packets.
What we try in this paper is not
to make all the $S$-matrices well-defined; rather, we ask
whether only well-defined $S$-matrices can be separated
from the remaining divergent ones {\it physically} or not.
And we ask whether the wave packet scheme is such a physical
method to separate the well-defined $S$-matrix subspace,
and whether the physical scheme justifies the formal schemes
within the subspace.

For field theory calculation,
one needs to specify a complete set for internal lines as well.
Unlike external lines,
we use plane wave complete set for simplicity.
One could use the wave packet set as well,
but the physical results will not depend on the choice of a complete set
for internal lines.
Despite the use of plane waves for internal lines,
the amplitude will still converge
if the plane wave couples with a wave packet.
This is because the wave packet suppresses the overlap of
wave functions (a precondition for scattering)
in the strong coupling region.

The spacetime action of the toy model is
\cite{callan,polchinski1}
\begin{equation}
S_{eff}=\int dX \, \sqrt{-G} e^{-2\Phi}
          \{ -\frac{1}{2} (\nabla T)^{2} + T^{2} - \frac{g}{3!} T^{3} \},
\end{equation}
where $dX = d^{2}X $.
The effective action would need higher order terms such as $T^{4}$,
to ensure exact conformal invariance of the $\sigma$-model.
But we regard the above action as a toy model
and truncate the action at $T^{3}$.
For the linear dilaton background, the action becomes
\begin{equation}
S_{eff}=\int dX -\frac{1}{2} \partial\varphi \cdot \partial\varphi
              - \frac{1}{2} (n^{2}-2) \varphi^{2}
              - \frac{g}{3!} e^{n \cdot X} \varphi^{3}
\end{equation}
after a field redefinition $\varphi=e^{n \cdot X}T$.
So, the field redefined tacyon $\varphi$
is massless as expected from $2D$ string vertex operator
(See the next section.).

The vertex Feynman rule is pathological in momentum space;
by a Fourier transform $\varphi_{k}=\int dX \,\varphi(X)e^{-ik \cdot X}$,
\begin{equation}
\int dX\, \frac{g}{3!} e^{n \cdot X} \varphi^{3}
   = \int \prod_{i=1}^{3} d\ki\,
       \varphi_{k_{i}}
       \int dX\,\frac{g}{3!}
       e^{i(k_{1}+k_{2}+k_{3}-in)\cdot X},
\end{equation}
which diverges since the three-point coupling diverges.
Here, $dk=d^{2}k/(2\pi)^{2}$.
Instead of expressing the integral in terms of a delta function formally,
we use wave packets in coordinate space.

For exercise, let us evaluate a couple of simple amplitudes.
The connected $S$-matrix from two wave packets to a plane wave is
\begin{equation}
iT(12 \rightarrow k)
    = \langle k|S-1|1,2 \rangle
    = ig \int dX \prod_{in} \Ti(X) e^{n \cdot X} e^{-ik \cdot X}.
                                            \label{eq:3pt}
\end{equation}
Note that this amplitude is explicitly finite
in spite of the use of a plane wave
because of the assumed form of $\Ti(X)$.
Similarly, the four-point amplitude is given by
\begin{equation}
T(12 \rightarrow 34)
     = g^{2} \int dX  dY\, e^{n\cdot (X+Y)}
       \prod_{in}\Ti(X) \prod_{out}\Ti^{*}(Y)
       \Delta(X-Y),
                                             \label{eq:4pt}
\end{equation}
where $\Delta(X-Y)$ is a plane wave propagator.
This amplitude is finite too,
since wave packets couple at each vertex.

It is straightforward to show the tree-level unitarity
for the four-point amplitude.
Using (\ref{eq:3pt}), (\ref{eq:4pt}),
and the Fourier representation of $\Delta(X-Y)$,
we get
\begin{eqnarray}
\sum_{k} T^{*}(34 \rightarrow k) T(12 \rightarrow k)
     & = & g^{2} \int dX  dY\, e^{n\cdot (X+Y)}
             \prod_{in}\Ti(X) \prod_{out}\Ti^{*}(Y) \nonumber \\
     &   & \times \int \dktilde\, e^{-ik\cdot (X-Y)}
\end{eqnarray}
and
\begin{eqnarray}
T(12 \rightarrow 34) - T^{*}(34 \rightarrow 12)
     & = & 2\pi ig^{2} \int dX  dY\, e^{n\cdot (X+Y)}
             \prod_{in}\Ti(X) \prod_{out}\Ti^{*}(Y) \nonumber \\
     &   & \times \int dk\, \delta^{2}(k^{2}) e^{-ik\cdot (X-Y)};
\end{eqnarray}
therefore, the unitary relation
$T_{fi} - T^{*}_{if} = i \sum_{n} T^{*}_{nf} T_{ni}$
holds.

\section{String Calculation}

The formal $S$-matrix at tree-level has been calculated
by a number of authors
in the context of the Liouville theory
\cite{gross,gupta}.
We first rederive the formal plane wave results
in order to compare the formal results with the wave packet results.

The $M$-tachyon amplitude is given by
the Polyakov path integral:
\begin{equation}
T(k_{1}, \cdots, k_{M})= e^{-\lambda\chi} \int [dX]\, e^{-S}
                         \prod_{i}^{M} V_{i}(k_{i}),
                                                    \label{eq:smatrix}
\end{equation}
where $S$ is given by (\ref{eq:action}) and
$V_{i}$ is the tachyon vertex operator given by
\begin{equation}
V_{i}(k_{i}) = g_{0} \int d^{2}\sigma_{i} \sqrt{g}
                         e^{ (ik_{i}+n) \cdot X(\sigma_{i}) }.
\end{equation}
The ghost path integral contribution is omitted
since it is factorized from the above path integral.
As usual, the mass shell condition follows from the requirement
that the vertex operator be a (1,1) tensor:
$k_{i}^{2} = 2-n^{2}$.

The path integral calculation is very similar to
the free bosonic one.
First, expand $X^{\mu}$
in eigenmodes of the laplacian, {\it i.e.}
\begin{equation}
X^{\mu}(\sigma)=cx_{0}^{\mu} + \sum_{m'} x_{m'}^{\mu} Y_{m'}(\sigma),
\end{equation}
where
\begin{eqnarray}
\nabla^{2} Y_{m} & = & -\omega_{m}^{2} Y_{m} \\
(Y_{m_{1}}, Y_{m_{2}}) & = &
                     \int d^{2}\sigma \sqrt{g}\, Y_{m_{1}} Y_{m_{2}}
                         = \delta_{m_{1} m_{2}},  \label{eq:norm}
\end{eqnarray}
where the sum over $m'$ denotes a nonzero mode sum.
We used that the zero mode, which satisfies
$\nabla^{2} Y_{0}=0$, is a constant, $Y_{0}=c$; $c$ is determined by
eq (\ref{eq:norm}).

The Green function on the sphere is defined in terms of these eigenmodes:
\begin{equation}
\prop{1}{2} = \sum_{m'} \frac{4\pi}{\omega_{m'}^{2}}
              Y_{m'}(\sigma_{1}) Y_{m'}(\sigma_{2}),
\end{equation}
which satisfies
\begin{equation}
-\frac{1}{4\pi} \nabla^{2} \prop{1}{2}
     = \frac{1}{\sqrt{g}} \delta^{2}(\sigma_{1}-\sigma_{2})-c^{2}.
\label{eq:poisson}
\end{equation}
The solution in conformal gauge is
\begin{equation}
G'(z_{1}, z_{2}) = -\ln | z_{1}-z_{2} |^{2}
                   + f(z_{1},\bar{z_{1}}) + f(z_{2},\bar{z_{2}}),
\label{eq:green}
\end{equation}
where $f$ arises from the $c^{2}$ term in (\ref{eq:poisson})
and depends on the world-sheet metric.

We will also need the renormalized Green function $G_{r}'(\sigma, \sigma)$;
define it as \cite{polchinski2}
\begin{equation}
G_{r}'(\sigma, \sigma)
     = \lim_{\sigma \rightarrow \sigma'} G'(\sigma, \sigma')
       + \ln d^{2}(\sigma, \sigma'),
\label{eq:renormed_green}
\end{equation}
where $d(\sigma, \sigma')$ is the geodesic distance between the two points.

Now, the amplitude (\ref{eq:smatrix}) becomes
\begin{eqnarray}
T & = & g_{0}^{M} e^{-\lambda\chi}
        \int \prod_{i} d^{2}\sigma_{i} \sqrt{g(\sigma_{i})}
        \int dx_{0}
        \exp\left[ icx_{0} \cdot \
            \left\{
                 \sum_{i=1}^{M}k_{i}-i(M\!-\!\chi)n
            \right\}
            \right]
  \nonumber \\
  &   & \times \prod_{m'} \int dx_{m'}
        \exp\left\{
          -\frac{\omega_{m'}^{2}}{8\pi}x^{\mu}_{m'}x_{\mu m'}
          + i x^{\mu}_{m'}
               \sum_{i=1}^{M}( k_{i}-in )_{\mu} Y_{m'}(\sigma_{i})
         \right.
  \nonumber  \\
  &   &  \left.
          -\frac{1}{4\pi} x^{\mu}_{m'} n_{\mu}
          \int d^{2}\sigma\sqrt{g}R Y_{m'}(\sigma)
         \right\} .
\end{eqnarray}
Note that the zero mode integral diverges.
We represent the integral as a delta function
with the complex argument for the time being. Then,
\begin{eqnarray}
T & = & g_{0}^{M} e^{-\lambda\chi}
        \left(\frac{2\pi}{c}\right)^{2}
        \delta^{2}\left( \sum_{i=1}k_{i}-i(M\!-\!\chi)n \right)
        \left(\mbox{det}' \frac{-\nabla^{2}}{8\pi^{2}}\right)^{-1}
  \nonumber \\
  &   & \times
        \int \prod_{i} d^{2}\sigma_{i} \sqrt{g(\sigma_{i})}
        \exp
        \left\{
         -\frac{1}{2}\sum_{i,j=1}^{M}(K_{i}\! \cdot\! K_{j})\prop{i}{j}
        \right.
  \nonumber \\
  &   & -\frac{i}{8\pi}\sum_{i}^{M} (K_{i}\! \cdot\! n)
                             \int d^{2}\sigma \sqrt{g} R \prop{}{j}
  \nonumber \\
  &   & \left.
         +\frac{1}{32\pi^{2}} n^{2}
            \int d^{2}\sigma d^{2}\sigma'
              \sqrt{g} R \sqrt{g'} R' G'(\sigma, \sigma') \right\},
\end{eqnarray}
where $\sqrt{g} R$ and its primed one are the functions of $\sigma$ and
$\sigma'$ respectively;
and $\Ki = \ki - in$.

In conformal gauge, the remaining integrals yields
the shifted Virasoro-Shapiro amplitude
\begin{eqnarray}
\lefteqn{ \left(\frac{1}{2}\right)^{M}
          \int \msr{i}{M}\, \mu(z_{i}) \pnorm{i<j}}
  \nonumber \\
  & = & \left(\frac{1}{2}\right)^{M}
       \int \msr{i=1}{M-3}\,
       |\zi|^{\expo{M-2}{i}} |1-\zi|^{\expo{M-1}{i}} \pnorm{i<j}
\label{eq:VS}
\end{eqnarray}
up to a constant,
where the Green functions
(\ref{eq:green}) and (\ref{eq:renormed_green}) are used.
$\mu(z_{i})$ is the standard Faddeev-Popov determinant
from {\it SL(2,C)} gauge fixing.
Also, we have used ``momentum conservation",
$\sum_{i} \ki = i(M\! - \!\chi)n$,
to drop the $f$-term in (\ref{eq:green})
and to obtain the second line of (\ref{eq:VS}).

Let us now turn to the corresponding wave packet calculation.
The path integrals can be carried out as before,
by changing the wave function in the vertex operator
$ e^{ik_{i}\cdot X(\sigma_{i})} $
to $T_{i}(X)$:
\begin{equation}
V_{i} = g_{0} \int d^{2}\sigma_{i} \sqrt{g}
                         e^{ n \cdot X(\sigma_{i}) } T_{i}(X).
\end{equation}
The nonzero mode integrals are not altered since the integrals are
well defined even for the plane wave case
because of the Gaussian nature of the integrals.
Therefore, the amplitude becomes
\begin{equation}
 T = N_{M} \int \dx
         \int \prod_{i} \dkitilde\,
                \rho_{i}\, e^{i\ki\cdot x_{0}} e^{(M-\chi)n\cdot x_{0}}
         \left\{ \int \msr{i}{M}\, \mu(z_{i}) \pnorm{i<j} \right\}
\label{eq:amplitude}
\end{equation}
with $N_{M}=(g_{0}/2)^{M} N$, where $N$ represents the various
irrelevant constants
(Here, it is understood that $\rho_{i}$ and
$e^{i\ki\cdot x_{0}}$
are appropriately complex conjugated for the out-states.
This notation is used henceforth.).
This is our main result.
If one takes the delta functions for $\rho_{i}$, {\it i.e.}
takes plane waves for the asymptotic states,
(\ref{eq:amplitude}) reproduces the formal result (\ref{eq:VS}).
In other words, the wave packet result is obtained
simply by inserting the wave packet weight $\rho_{i}$
and the momentum integral
after the formal plane wave calculations.
In this way, the wave packet scheme justifies the formal results
provided the integral (\ref{eq:amplitude}) is really finite.

Although this result itself may not be surprising,
this is not a trivial result.
If one wants to avoid formal approaches,
one can no longer use momentum conservation rule.
But it was essential to use the rule in plane wave calculation
in order to drop the $f$-term in the Green function
and to cancel the Faddeev-Popov determinant.
Without the help of the conservation rule, how can one derive
(\ref{eq:amplitude})?
The point is that one can use the conservation rule in a sense,
even though the momenta $k_{i}$ appearing in (\ref{eq:wavepacket})
are actually real by definition.
For instance, the following integral vanishes
without the help of the formal conservation rule;
\begin{eqnarray}
\lefteqn{ \int \dx \int \prod_{i} \dkitilde\,
                \rho_{i}\, e^{i\ki\cdot x_{0}} e^{(M-\chi)n\cdot x_{0}}
                \left\{
                 \sum_{l} ik_{l}^{\nu} +(M\!-\!\chi)n^{\nu} \right\} }
         \nonumber \\
  & = & \int \prod_{\mu \neq \nu} dx_{0}^{\mu}
           \left.
             T_{i}(x_{0})\, e^{(M-\chi)n\cdot x_{0}}
           \right|_{x_{0}^{\nu}=-\infty}^{\infty} = 0
\end{eqnarray}
since the wave functions $T_{i}(x)$ fall off faster than the exponential.
Similarly, one can show that for any analytic function $f(x)$
which admits a Taylor expansion in $x$,
\begin{eqnarray}
\lefteqn{ \int \dx \int \prod_{i} \dkitilde\,
                \rho_{i}\, e^{i\ki\cdot x_{0}} e^{(M-\chi)n\cdot x_{0}}
                f\left( \sum_{l} ik_{l} + (M\!-\!\chi)n \right) }
        \nonumber \\
  & = & \int \dx \int \prod_{i} \dkitilde\,
                \rho_{i}\, e^{i\ki\cdot x_{0}}
			   e^{(M-\chi)n\cdot x_{0}} f(0).
\label{eq:mom_conserv}
\end{eqnarray}

We can show the tree-level unitarity
of the four-tachyon amplitude using (\ref{eq:amplitude})
for the lowest pole (tachyon).
{}From (\ref{eq:amplitude}),
the process from two wave packets to a tachyon plane wave
gives the following transition amplitude:
\begin{equation}
T(12 \rightarrow k) = N_{3} \int \dx\,
         e^{n\cdot x_{0}} \prod_{in} T_{i}(x_{0}) e^{-ik\cdot x_{0}}.
\end{equation}
Note that the amplitude is finite
and the result agrees with the field theory result (\ref{eq:3pt}).
Then,
\begin{eqnarray}
\lefteqn{ \sum_{k} T^{*}(34 \rightarrow k) T(12 \rightarrow k) }
		\nonumber \\
    & = &  N_{3}^{2}\int \dx  \dy\, e^{n\cdot (x_{0}+y_{0})}
             T_{1}(x_{0}) T_{2}(x_{0})
             T_{3}^{*}(y_{0}) T_{4}^{*}(y_{0})
             \int \dktilde\, e^{-ik\cdot (x_{0}-y_{0})}.
\label{eq:sumTT}
\end{eqnarray}
For the four-point amplitude,
it is more convenient to rewrite the amplitude
by inserting a delta function for the comparison
with (\ref{eq:sumTT}):
\begin{eqnarray}
 T(12 \rightarrow 34)
    & = & N_{4} \int \dx \dy
          \int dk \prod_{i} \dkitilde\,
                \rho_{i}\, e^{i(k_{1}+k_{2}-k-in)\cdot x_{0}}
                         e^{i(-k_{3}-k_{4}+k-in)\cdot y_{0}}
    \nonumber \\
    &   & \times \int d^{2}z\,
                |z|^{\expo{1}{2}} |1-z|^{\expo{1}{3}}.
\end{eqnarray}
Because $|z|^{\expo{1}{2}}$ is analytic in $(K_{1}\!\cdot\! K_{2})$,
one can use (\ref{eq:mom_conserv}) and obtain
\begin{eqnarray}
T(12 \rightarrow 34)
    & = & N_{4} \int \dx \dy
          \int dk \prod_{i} \dkitilde\,
                \rho_{i}\, e^{i(k_{1}+k_{2}-k-in)\cdot x_{0}}
                         e^{i(-k_{3}-k_{4}+k-in)\cdot y_{0}}
    \nonumber \\
    &   & \times \int d^{2}z\,
                |z|^{k^{2}-4+n^{2}-i\epsilon} |1-\zi|^{\expo{1}{3}}.
\end{eqnarray}
The above amplitude contains the tachyon pole at
$k^{2}-3+n^{2}=-1$;
\begin{eqnarray}
\lefteqn{ T(12 \rightarrow 34) - T^{*}(34 \rightarrow 12)}
    \nonumber \\
    & = & 4\pi^{2} i N_{4} \int \dx  \dy\, e^{n\cdot (x_{0}+y_{0})}
             T_{1}(x_{0}) T_{2}(x_{0})
             T_{3}^{*}(y_{0}) T_{4}^{*}(y_{0})
    \nonumber \\
    &   & \times \int dk\, \delta(k^{2}+n^{2}-2)\,
		e^{-ik\cdot (x_{0}-y_{0})}
          + \mbox{analytic at $k^{2}=2-n^{2}$},
\end{eqnarray}
which is proportional to (\ref{eq:sumTT}).
Finally, unitary relation determines
the normalization of the vertex operator, $g_{0}$,
in terms of the various constants and the string coupling
$e^{-\lambda\chi}$ \cite{weinberg},
and this completes the proof of unitarity for the lowest pole.

\section{Revival of Divergence and Comment}

Although three-, four-, and five-point amplitudes are finite
in the wave packet scheme,
this is not true for the amplitudes with more external legs.

This is easily seen in effective field theory.
Figure 1 shows a six-point diagram which diverges.
The diagram contains the $W$-vertex
at which no wave packet couples;
therefore, the vertex integral diverges
like plane wave three-point amplitude.

However, needless to say,
the problem is not caused because plane waves
are used for the internal lines;
any complete set should give the same answer
(even the set of well localized wave packets).
Rather, the problem is caused by the facts
(1) that the $S$-matrix formalism obtains the amplitudes
from $t=-\infty$ to $t=\infty$ and
(2) that momentum conservation does not hold for this background.
Due to these properties,
even if we confine asymptotic states in the weakly coupled region,
intermediate states can still travel to the strongly coupled region
(Figure 2).
The trip to the strongly coupled region is the origin
of the divergent answer.

Although the divergence is not easily seen in (\ref{eq:amplitude}),
the divergence should exist in string calculation as well,
because low-energy Feynman diagrams are reproduced
by taking limits of moduli space integrals
in string amplitudes.

We finish with the following remark
about another aspect of the wave packet scheme.
As is well-known,
the $S$-matrix formalism is the only formalism useful in string theory.
However, the formalism is basically a global formalism,
because the formalism gives amplitudes
from $t=-\infty$ to $t=\infty$
and plane waves are used for the asymptotic states.
This global property makes local physics obscure
when one is interested in spacetime dependence in string theory.
The use of wave packets in string theory, therefore,
may help to extract local physics even in the $S$-matrix formalism.

\vspace{.1in}

\begin{center}
    {\Large {\bf Acknowledgements} }
\end{center}
\vspace{.1in}

I am grateful to Joe Polchinski for his continuous help
and interest throughout the work.
I am also pleased to thank John LaChapelle for his comments on the draft.
This research was supported in part by the Robert A. Welch
Foundation, NSF Grant PHY 8904035 and 9009850.

\pagebreak

\end{document}